\documentclass[sigconf]{acmart}
 \acmConference[EASE 2026]{The 30th International Conference on Evaluation and Assessment in Software Engineering}{9–12 June, 2026}{Glasgow, Scotland, United Kingdom}
\usepackage{multirow}
\usepackage{array} 
\usepackage{graphicx}
\usepackage{subcaption}
\usepackage{xspace}
\usepackage{mdframed}
\usepackage{multirow}
\usepackage{adjustbox}
\usepackage{breakurl}
\usepackage{url}
\usepackage{graphicx}
\usepackage{booktabs}
\usepackage{threeparttable}
\usepackage{tabularx}
\usepackage{multirow}
\usepackage{subcaption}
\usepackage{hyperref}
\usepackage{xurl} 
\usepackage{etoolbox}
\newcommand{\name}
{\textit{SelfHeal}\xspace}

\newcommand{\benchmark}%
{\textit{AgentDefect}\xspace}

\usepackage{caption} 
\usepackage{caption} 
\AtBeginEnvironment{mdframed}{\vspace{1em}}
\AfterEndEnvironment{mdframed}{\vspace{1em}}

\newcommand{\etal}{\textit{et al.}\xspace}

\AtBeginDocument{%
  }

\setcopyright{acmlicensed}
\copyrightyear{2018}
\acmYear{2026}
\acmDOI{XXXXXXX.XXXXXXX}

\acmConference[EASE]{30th International Conference on Evaluation and Assessment in Software Engineering (EASE 2026)}{June 09--12,
  2026}{Glasgow, United Kingdom}

\acmISBN{978-1-4503-XXXX-X/2018/06}

\begin{document}

\title{SelfHeal: Empirical Fix Pattern Analysis and Bug Repair in LLM Agents}

\author{Niful Islam, Muhammad Anas Raza, Mohammad Wardat}
\authornote{Corresponding author. Email: wardat@oakland.edu}
\affiliation{%
    \department{Department of Computer Science and Engineering}
    \institution{Oakland University}
    \city{Rochester}
    \state{Michigan}
    \country{USA}
}
\email{{islam3,mraza,wardat}@oakland.edu}





\begin{abstract}
Large Language Models (LLMs) have transformed software development and AI applications. While LLMs are designed for text processing, LLM agents extend this capability by enabling autonomous actions, tool use, and multi-step task completion. As this field grows, developers face new challenges in debugging these complex systems. To address this challenge, we present the first empirical study on bug fix patterns in LLM agents. We study buggy posts and code snippets from three platforms: Stack Overflow, GitHub, and HuggingFace Forums. We examine their fix patterns, the components where fixes are applied, and the programming languages and frameworks involved. Furthermore, we introduce \benchmark, the first benchmark dataset for bugs in LLM agents. The dataset contains 37 runtime buggy instances along with fixed code and test files. Finally, we present \name, a multi-agent system designed to fix bugs in LLM agents. The system leverages two independent ReAct agents: the fix agent and the critic agent. These agents use tools that provide both internal knowledge (fix rules) and external knowledge (web search) to propose and validate fixes. Our evaluation shows that \name with Gemini 3 Pro as the backbone LLM outperforms both baseline and state-of-the-art approaches by a significant margin.
\end{abstract}



\maketitle
\thispagestyle{plain}
\pagestyle{plain}

\section{Introduction}
\label{sec:intro}

With improvements in Deep Learning (DL) and Large Language Models (LLMs), their applications span diverse disciplines, including healthcare \cite{rousmaniere2026large}, software engineering \cite{raza2025graph}, and cybersecurity \cite{islam2025anomaly}. In software engineering, there is an increasing shift toward Agentic AI, in which autonomous agents perform complex tasks such as reasoning and tool manipulation \cite{chowa2026language}. This shift introduces unique hurdles that current engineering practices are still evolving to address, primarily due to the inherent non-determinism of these systems. AI agents are composed of multiple tightly coupled components, including reasoning modules, memory mechanisms, planning strategies, and external tool interfaces \cite{chowa2026language}, where errors in any component can propagate across the entire execution pipeline, leading to cascading failures. This interdependence, combined with the black box and non-deterministic nature of LLMs, makes debugging substantially more challenging than traditional software, as faults are often non-local, temporally distributed, and difficult to reproduce.

Furthermore, the ecosystem remains in a nascent stage. It is characterized by new frameworks with small communities and a lack of standardized taxonomies or comprehensive documentation. New frameworks related to agents and their versions are released every month. For instance, LangChain, the most popular library for constructing LLM agents, was released in late 2022 and has released over 45 new versions in 2025 and over 65 versions in 2024 \cite{langchain-pypi}. LlamaIndex, the second most popular library, released over 60 versions in 2025 and over 140 versions in 2024 \cite{llamaindex-pypi}. Each version introduces new features and changes. Therefore, developers often fail to adapt to the new changes and receive less community support since fewer developers are involved in the field.

Prior research has established various taxonomies for software bugs, ranging from general software systems \cite{catolino2019not} to specific deep learning artifacts and attention-based mechanisms \cite{islam2019comprehensive, jahan2025taxonomy}. While recent studies have begun to evaluate the performance of agents on coding benchmarks and trace their autonomous workflows \cite{yang2024swe, deshpande2025trail}, these evaluations primarily focus on agent effectiveness rather than agent correctness. To date, no empirical study has systematically investigated fix patterns that can be used to fix bugs within agentic systems. Furthermore, although several frameworks have been proposed that employ agent-based architectures for automated bug fixing in conventional software systems \cite{yang2024swe} \cite{zhang2024autocoderover} \cite{xia2025demystifying}, these approaches treat agents as reliable repair tools rather than faulty software artifacts. As a result, no dedicated framework exists for diagnosing and repairing bugs specific to agentic systems. Although initial efforts attempt to define agent defects by analyzing discrepancies between developer logic and generated content \cite{ning2024defining}, the literature still lacks a comprehensive methodology for fixing real-world agent bugs reported by practitioners across platforms such as community forums, version control repositories, and model hubs.

To address these gaps, we present the first comprehensive study on bug fix patterns in LLM agents. Following prior works \cite{islam2026agentsfail, islam2019comprehensive}, we have collected posts and code snippets from Stack Overflow, GitHub commits, GitHub issues, and HuggingFace Forums, and analyzed the bug fix patterns within them. In addition to the fix pattern, our analysis also includes the following dimensions: the specific component of the LLM agent where each fix was applied, the programming languages and frameworks used, the rationale behind selecting each fix pattern, and the buggy and fixed code patches (where provided by users). When fixes involved framework version issues, we also documented the buggy and corrected version requirements. Secondly, we curated the first benchmark dataset on runtime bugs in LLM agents, \benchmark, containing 37 executable buggy code samples. Each instance includes the buggy code, the user's one-sentence intent, the corresponding fixed code, a test file, a readme file explaining execution instructions, and a requirements file with library versions. Additionally, we present \name, a comprehensive framework for analyzing and repairing bugs in agentic systems. \name employs two independent ReAct agents: a fix agent that generates repairs and a critic agent that validates them. Both agents are empowered by internal and external knowledge sources. Through careful design, \name outperforms both baseline and state-of-the-art (SoTA) approaches. Overall, the primary contributions of the research works are as follows:

\begin{itemize}
    \item We have conducted the first empirical study on bug fix patterns in LLM-based agents. We also provide the curated dataset with code patches and component-level annotations that can be used by future studies.
    \item We present \benchmark, the first benchmark dataset of runtime bugs in LLM agents. The dataset contains 37 executable instances, each with buggy and fixed code, user intent, test files, execution instructions, and versioned dependencies.
    \item We present \name, a multi-agent system for fixing bugs in agentic systems. It combines a fix agent and a critic agent, both empowered by internal and external knowledge sources, to iteratively repair and validate buggy code.
    \item Our proposed solution significantly outperforms baseline and SoTA approaches on the newly developed dataset.
\end{itemize}
\vspace{-7pt}

\section{Empirical Study}
\label{sec:empirical_study}
We followed previous studies on LLM agents \cite{islam2026agentsfail}, to collect and annotate the dataset for our empirical study. The data collection and annotation process is explained in the subsequent sections. 
\vspace{-5pt}

\subsection{Data Collection}
Following prior studies \cite{islam2019comprehensive, islam2026agentsfail}, we sourced our dataset from three platforms, namely Stack Overflow, GitHub (issues and commits), and HuggingFace Forums. For Stack Overflow we searched for posts containing keywords such as LangChain, LlamaIndex, Semantic Kernel, CrewAI, AutoGen, OpenAI, and Ollama and found 2,734 instances. We manually removed irrelevant posts (i.e., posts that do not discuss LLM agents), posts with insufficient descriptions, and posts without a determined solution that was either verified in the post or publicly available, and retained 665 unique annotated Stack Overflow posts. For GitHub and HuggingFace Forums, we applied the same process described in \cite{islam2026agentsfail} and obtained 51 instances from GitHub issues, 129 from GitHub commits, and 85 from HuggingFace Forums with no duplicate entries. We studied a total of 930 unique buggy instances from the three platforms and analyzed their fix patterns.
\vspace{-5pt}

\subsection{Data Annotation}
We began the annotation process using fix patterns from existing taxonomies of software and data science bugs \cite{chen2025towards, pan2009toward}. After studying these taxonomies, we adopted fix patterns that were present in LLM agent-related issues, while patterns not observed in our dataset were excluded. We then annotated the fix patterns in our dataset and identified eight patterns not covered by prior studies. For these newly identified patterns, we derived labels using an open card sorting method, which we added for labeling. Definitions of all fix patterns are provided in Section \ref{sec:data-descriptions}. In addition to fix patterns, we annotated the component where the fixes are applied, following the four-component framework for LLM agents described by prior works \cite{ning2024defining, islam2026agentsfail}. We also recorded the programming language and framework used to build each agent. For code-related issues, we documented the buggy and fixed source code along with output or error messages. For library version issues, we recorded the buggy and fixed requirements when the buggy version is specified. Each annotation included a one-line rationale explaining the fix pattern selection and cited external resources when solutions were found outside the post. Notably, to identify bug fixes and their patterns, we analyzed verified answers and replies from Stack Overflow and HuggingFace Forums, before-and-after code changes from GitHub commits, and associated pull requests for GitHub issues. When posts in Stack Overflow or HuggingFace Forums lacked verified answers, we searched for fixes from external sources and documented these external resources.

For the data annotation, we followed previous literature \cite{islam2019comprehensive} where two PhD students independently labeled fix patterns and components. We measured inter-annotator agreement using Cohen's kappa coefficient after every 10\% of the Stack Overflow dataset was annotated. When annotators disagreed, an expert assisted in reaching a consensus. Programming languages and frameworks were determined from Stack Overflow tags or GitHub repository metadata and annotated collaboratively. For the smaller GitHub and Huggingface datasets, agreement was measured after completing the full annotation. Final kappa scores for the Stack Overflow dataset were 0.983 for fix patterns and 0.808 for components. For GitHub commits, scores were 0.953 and 0.919; for GitHub issues, 1.0 and 0.812; and for HuggingFace forums, 0.973 and 0.973 for fix patterns and components, respectively. The complete table of kappa scores after each interval is provided in the supplementary material \cite{selfheal2026}. When multiple fix patterns were required to resolve a single bug, we created separate entries for each pattern. This resulted in 89 additional entries in the Stack Overflow dataset, 5 in the GitHub commits dataset, and 1 in the HuggingFace forums dataset. In total, it resulted in 1025 entries for all the datasets combined. 
\vspace{-5pt}
\subsection{Description of Bug Fix Patterns}
\label{sec:data-descriptions}
To systematically evaluate the repair capabilities of various agentic systems, we categorize observed fixes into a taxonomy of 23 distinct patterns. Fix patterns documented in prior literature on traditional software systems were adopted and cited accordingly \cite{chen2025towards} \cite{pan2009toward}, while patterns not identified in existing work were excluded from the taxonomy. The description of each fix pattern is listed below.

\subsubsection{Add New Attribute (ANA)}
Augments a function or method call by introducing explicit arguments that replace insufficient default values or satisfy previously unmet functional requirements. This pattern typically arises when implicit assumptions encoded in defaults fail under specific execution contexts, requiring the caller to pass configuration parameters directly~\cite{chen2025towards} (e.g., \cite{stackoverflowTryingCreate}).

\subsubsection{Remove Attribute (RA)}
Augments a function or method call by removing explicit arguments so the call uses default values or avoids unsupported parameters ~\cite{chen2025towards} (e.g., \cite{githubLLamaSharpEmbeddingsException}).
\vspace{-4pt}

\subsubsection{Addition of Precondition Check (AOPC)}
Implements defensive programming constructs by inserting conditional checks that validate system state, environmental assumptions, or input invariants before execution proceeds. This prevents invalid states from propagating deeper into the program and reduces the likelihood of runtime failures~\cite{pan2009toward} (e.g.,~\cite{stackoverflowTryingCreate}).
\vspace{-4pt}
\subsubsection{Change Version (CV)} Modifies dependency or package version specifications to restore compatibility between interacting libraries or frameworks. This pattern addresses failures caused by deprecated APIs, breaking changes, or version-specific regressions that emerge during library evolution~\cite{ringer2021proof} (e.g., \cite{githubRunningStop}).
\vspace{-4pt}
\subsubsection{Install Library (IL)} Resolves execution failures originating from missing external dependencies by explicitly installing or declaring required libraries. This pattern reflects environment-level repairs rather than code-level logic changes and is common in deployment or reproduction scenarios~\cite{jia2021symptoms} (e.g., \cite{githubBUGModule})
\vspace{-4pt}
\subsubsection{Addition of Operations (AOO)} Introduces previously omitted functional steps into the execution pipeline, such as data loading, preprocessing, or intermediate transformation stages. In complex systems like RAG architectures, this often involves inserting embedding generation or retrieval components necessary for downstream tasks~\cite{pan2009toward} (e.g., \cite{stackoverflowEmbeddingDocuments}).
\vspace{-4pt}
\subsubsection{Removal of Operations (ROO)} Eliminates obsolete, redundant, or logically incorrect code segments that contribute unnecessary computation or interfere with correct execution. This pattern streamlines control flow and can improve both correctness and efficiency~\cite{pan2009toward} (e.g., \cite{githubFixSpan}).
\vspace{-4pt}
\subsubsection{Change Data Type (CDT)} Corrects mismatches between expected and actual data types in variable assignments or function calls. By aligning data representations with API or language constraints, this pattern prevents type-related runtime exceptions and semantic errors~\cite{chen2025towards} (e.g., \cite{githubGenerateLlama32}).
\vspace{-4pt}
\subsubsection{Change Prompt (CP)} Refines the structure, phrasing, or formatting of prompts provided to Large Language Models in order to improve output relevance, consistency, or task alignment. This includes modifying instruction wording or adding clarifying context (e.g., \cite{githubChessSample}).
\vspace{-4pt}
\subsubsection{Fix Syntax (FS)} Addresses violations of programming language grammar, including missing delimiters, incorrect indentation, or malformed expressions. These fixes are necessary to restore parsability and enable successful compilation or interpretation~\cite{chen2025towards} (
e.g., \cite{githubHotfixRemove}).
\vspace{-4pt}
\subsubsection{Change Reference (CR)} Fixes incorrect references to internal modules, classes, or external namespaces. Such errors typically arise from refactoring, file relocation, or inconsistent naming and result in unresolved symbols at runtime or import time. Due to rapid developments in the field, modules often relocate to different packages. This requires developers to update their import statements (e.g., \cite{stackoverflowModuleNotFoundErrorModule}).
\vspace{-4pt}
\subsubsection{Use Different Module (UDMo)} Replaces an existing software component with an alternative module that provides equivalent functionality but improved compatibility, stability, or performance for the target task. This pattern reflects higher-level design substitution rather than localized fixes (e.g., \cite{githubPfdreaderMt7180quaiglec29f047}).
\vspace{-4pt}
\subsubsection{Change External Resources (CER)} Reconfigures dependencies on external resources, including API endpoints, authentication credentials, environment variables, or remote services. These changes are not related to the agents (e.g., \cite{stackoverflowUsingVicuna}).
\vspace{-3pt}
\subsubsection{Change Function (CF)}Replaces an incorrect function with the correct one that provides the intended functionality. This occurs when developers initially select the wrong function or API for a specific task.~\cite{pan2009toward} (e.g., \cite{githubFixrequires_dependencies}).
\vspace{-4pt}
\subsubsection{Add Exception Handling (AEH)} Encapsulates error-prone code regions within try-except or equivalent constructs to improve system robustness. This pattern allows failures to be caught, logged, or recovered from gracefully instead of terminating execution abruptly~\cite{pan2009toward} (e.g., \cite{stackoverflowLangChainQuerying}).
\vspace{-4pt}
\subsubsection{Fix Attribute Name (FAN)}: Resolves inconsistencies or typographical errors in attribute or parameter names so that they align with the definitions provided by the underlying API or object interface. These fixes are often subtle but critical for correct binding~\cite{chen2025towards}(e.g., \cite{githubWithEmpty}).
\vspace{-4pt}
\subsubsection{Change Parameter Value (CPV)}: Updates literal constants or variable values passed to functions to better match expected operational ranges or logic constraints. Unlike data type changes, this pattern preserves type correctness while adjusting semantics~\cite{chen2025towards} (e.g., \cite{githubFixUpdate}).
\vspace{-4pt}
\subsubsection{Change Input Data (CID)}: Cleans, restructures, or reformats malformed or incompatible external data to ensure that downstream components can process it correctly. This pattern focuses on data integrity rather than code modification (e.g.,\cite{github23LittleLittleCloudAgentChatRooma8fe2e4}).
\vspace{-4pt}
\subsubsection{Add Input Data (AID)}: Resolves invocation or execution failures by supplying required input data that was previously missing or implicitly assumed. This includes mandatory configuration files, parameters, or runtime artifacts~\cite{endres2019infix}.
\vspace{-4pt}
\subsubsection{Use Different Model (UDM)} Enables a model-level substitution, such as switching to a different LLM, to satisfy constraints related to performance, cost, availability, or compatibility. This pattern reflects strategic adaptation at the system level (e.g., \cite{githubFixedGpt4omini}).
\vspace{-4pt}
\subsubsection{Change Parameter Order (CPO)} Reorders arguments in function calls to align with the expected positional signature, preventing logical errors caused by incorrect argument binding~\cite{yang2024morepair}.

\vspace{-4pt}
\subsubsection{Move Code to Different Scope (MCTDS)} Corrects errors caused by incorrect indentation or improper code placement within control flow structures (e.g., \cite{githubProcessEvent}).
\vspace{-4pt}
\subsubsection{Fix Data Access (FDA)} Corrects the logic governing the retrieval, indexing, or slicing of structured output data (e.g., 
\cite{githubMergePull}).

\begin{figure*}[htbp]
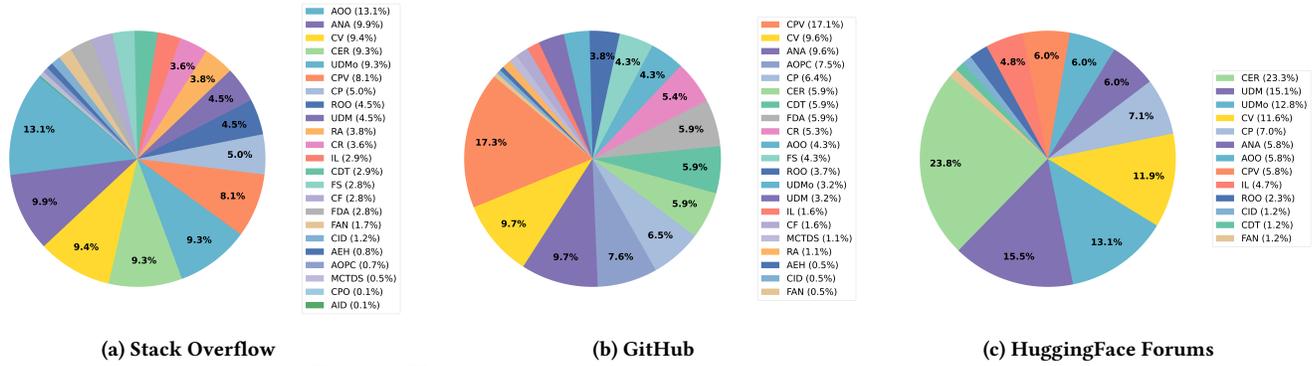

    \centering
    \begin{subfigure}[b]{0.32\textwidth}
        \centering
        \includegraphics[width=\linewidth]{SO_Fix_Pattern.png}
        \caption{Stack Overflow}
        \label{fig:fix-pattern-distribution-so}
    \end{subfigure}
    \hfill 
    \begin{subfigure}[b]{0.32\textwidth}
        \centering
        \includegraphics[width=\linewidth]{GH_Fix_Pattern.png}
        \caption{GitHub}
        \label{fig:fix-pattern-distribution-gh}
    \end{subfigure}
    \hfill 
    \begin{subfigure}[b]{0.32\textwidth}
        \centering
        \includegraphics[width=\linewidth]{HF_Fix_Pattern.png}
        \caption{HuggingFace Forums}
        \label{fig:fix-pattern-distribution-hf}
    \end{subfigure}
    
    \caption{Fix pattern distribution across different data sources.}
    \label{fig:fix-pattern-distribution}
\end{figure*}
\vspace{-3pt}

\subsection{Empirical Analysis}

\vspace{-3pt}
Our empirical analysis is conducted to answer the following research question.
\begin{figure*}[htbp]
    \centering
    \begin{subfigure}[b]{0.32\textwidth}
        \centering
        \includegraphics[width=\linewidth]{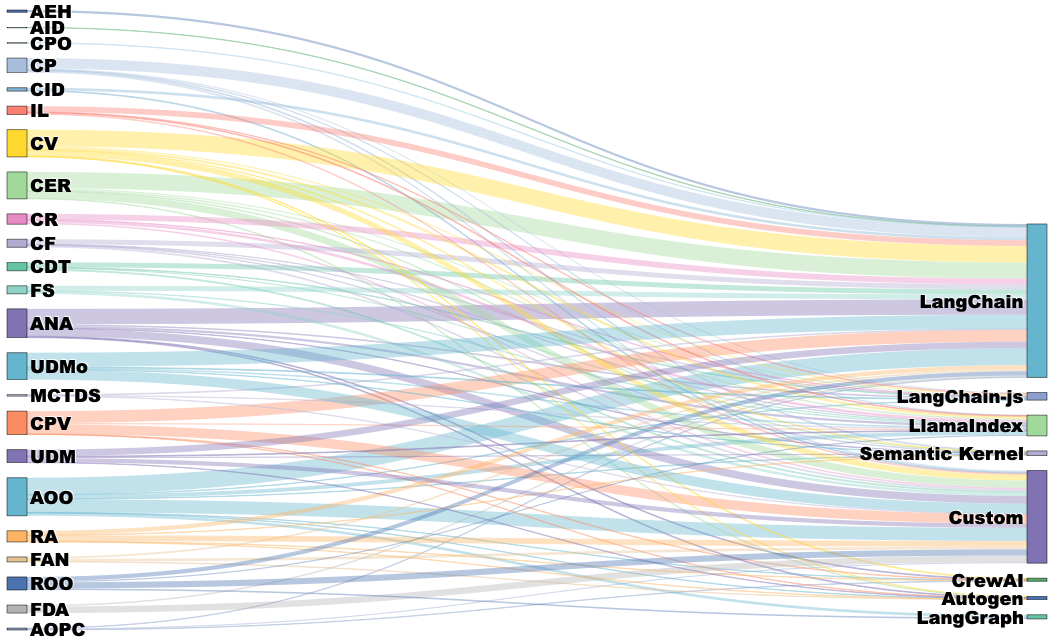}
        \caption{Stack Overflow}
        \label{fig:fix-pattern-distribution-framework-so}
    \end{subfigure}
    \hfill 
    \begin{subfigure}[b]{0.32\textwidth}
        \centering
        \includegraphics[width=\linewidth]{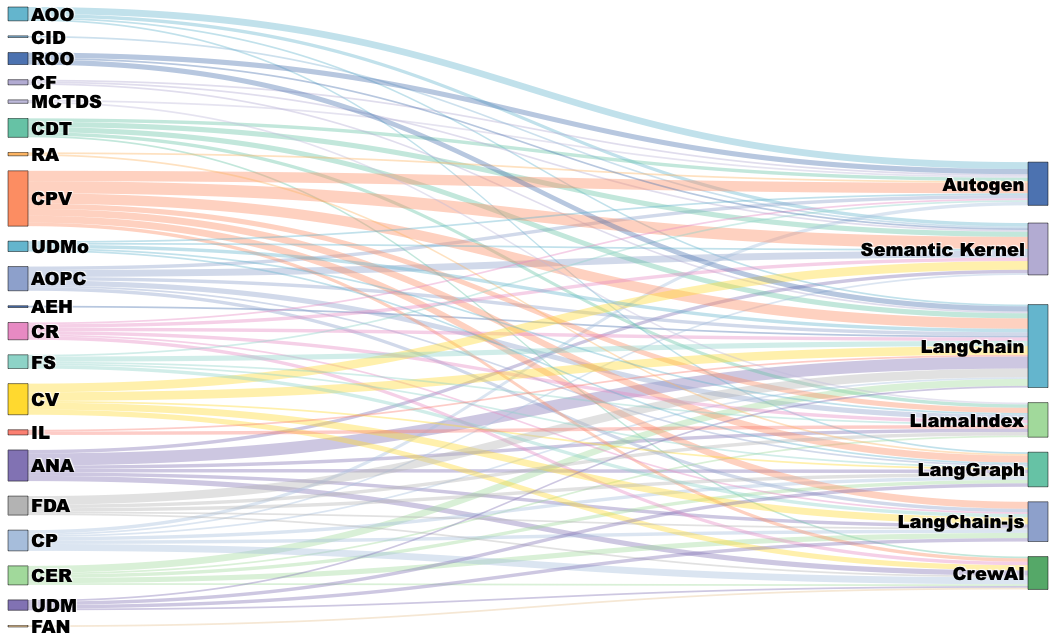}
        \caption{GitHub}
        \label{fig:fix-pattern-distribution-framework-gh}
    \end{subfigure}
    \hfill
    \begin{subfigure}[b]{0.32\textwidth}
        \centering
        \includegraphics[width=\linewidth]{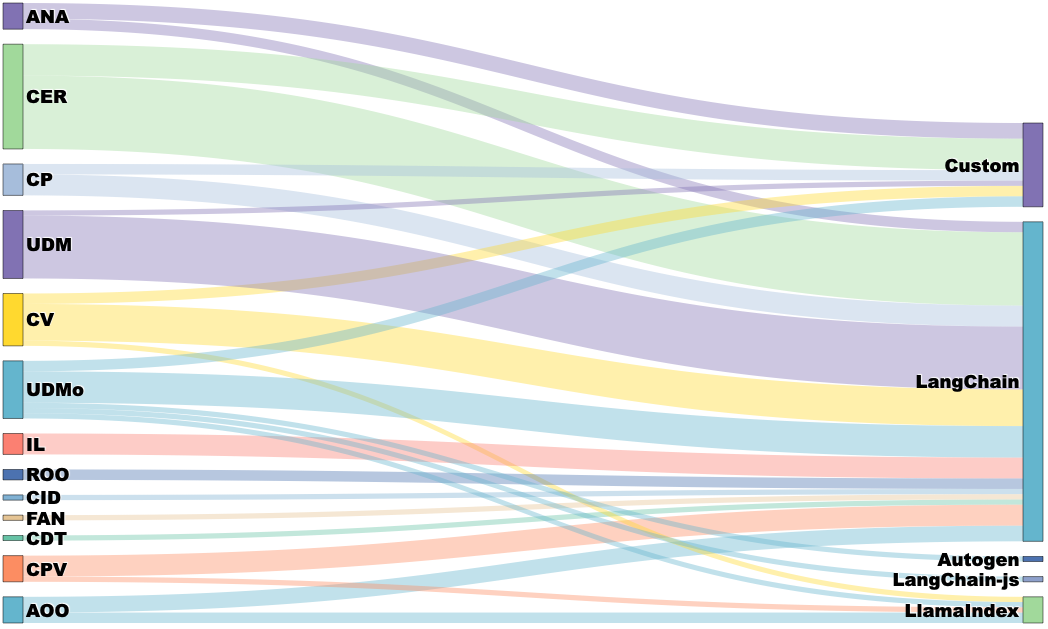}
        \caption{HuggingFace Forums}
        \label{fig:fix-pattern-distribution-framework-hf}
    \end{subfigure}
    
    \caption{Fix pattern distribution across frameworks.}
    \label{fig:fix-pattern-distribution-framework}
\end{figure*}

\vspace{-3pt}
\begin{itemize}
    \item \textbf{RQ1 (Frequency):} What are the common fix patterns in LLM agents, and how frequently do they occur?
    \item \textbf{RQ2 (Variation):} Do the fix patterns in agentic systems and their distributions differ from those in traditional software systems?
    \item \textbf{RQ3 (Spread):} What is the distribution of fix patterns across different libraries?
    \item \textbf{RQ4 (Challenges):} What are the challenges in fixing bugs in LLM agents? 
\end{itemize}

\vspace{-3pt}
\subsubsection{Frequency:}
Figure \ref{fig:fix-pattern-distribution} presents the distribution of fix patterns across Stack Overflow, GitHub, and HuggingFace Forums data sources. As shown in the figure, Addition of Operations (AOO) is the most common fix pattern on Stack Overflow, Change Parameter Value is the most common on GitHub, and Change External Resources is the most frequent on HuggingFace Forums. These differences stem from the nature of each platform. On GitHub, agents are typically already developed and require parameter tuning for optimization or adaptation to specific use cases. On HuggingFace Forums, the prevalence of Change External Resources reflects the fact that HuggingFace Hub provides LLMs for use in agentic systems. Since these models must be downloaded locally, they introduce additional complexity in configuring external resources. Therefore, users prefer to discuss these problems in HuggingFace Forums with the specialized community there. Nevertheless, the high frequency of AOO and ANA type fix patterns in the Stack Overflow dataset indicates that many bugs arise from missing or mishandled operations. This highlights the need for more comprehensive documentation and best practices in agent development. Additionally, the higher number of fix patterns related to version change (CV) indicates the rapid development in agentic libraries, which forces developers to frequently update their code to maintain compatibility, a trend also found in the early stages of DNN development \cite{islam2020repairing}. Lastly, the lower frequency of data-related fix patterns (e.g., AID) compared to those in DNN \cite{islam2020repairing} is due to the fact that building agents generally do not require fine-tuning LLMs. Rather, agents use pretrained LLMs, which makes them less dependent on input data.

\vspace{-8pt}

\begin{mdframed}[backgroundcolor=gray!5]
Addition of Operations is most common in Stack Overflow, Change Parameter Value dominates GitHub, and Change External Resources is most frequent in HuggingFace Forums.
\end{mdframed}
\vspace{-8pt}

\subsubsection{Variation:}
The distribution of fix patterns in agentic systems diverges fundamentally from established paradigms in traditional software engineering and deep learning by shifting from deterministic logic and structural optimization toward semantic orchestration and workflow augmentation. While traditional fixes in software systems are primarily characterized by repairs to deterministic logic, such as implementing precondition checks \cite{pan2009toward}, resolving system interaction bugs involving file management \cite{liu2024mining}, or implementing path resolution protocols \cite{liu2024mining}, and DL systems focus on data-centric adjustments, most notably Data Dimension (18.8\% in Stack Overflow) \cite{islam2020repairing} and Network Connection (17.8\% in Stack Overflow) \cite{islam2020repairing} to ensure model convergence, agentic AI introduces a unique class of behavior-centric repairs. As evidenced by empirical data, the high prevalence of Addition of Operations (13.1\% in Stack Overflow) and Add New Attribute (9.9\%) indicates that agentic failures often stem from missing functional components or inadequate tool-calling protocols rather than broken syntax or tensor misalignment.

This represents a significant departure from DL fix patterns where logic is embedded within weights and architecture \cite{islam2020repairing}; in agentic systems, the logic resides in the `Agentic Flow,' where swapping a library via Use Different Module is often a more frequent solution than swapping the core LLM engine. Furthermore, the emergence of Change External Resources as a dominant pattern in specialized forums (23.3\% in HuggingFace) highlights an environmental dependency unique to agents that must dynamically manage local model hosting and API continuity. This complexity mirrors the high maintenance costs found in early DNN development due to library versioning, where versioning-related fixes reached 17.6\% in GitHub \cite{islam2020repairing}, but introduces a distinct layer of orchestrational autonomy and cross-platform interaction challenges \cite{liu2024mining}. Lastly, although prompts play a critical role in LLM based systems and changing prompts represents a unique fix pattern for agentic systems, our study indicates that only a small number of fixes (5-7\%) require prompt modification.

\begin{mdframed}[backgroundcolor=gray!5] The distribution of fix patterns in agentic systems diverges from traditional and deep learning paradigms by prioritizing semantic orchestration and workflow augmentation over deterministic logic and structural optimization. 
\vspace{-0.1cm}
\end{mdframed}
\vspace{-0.2cm}

\subsubsection{Spread:}
Figure \ref{fig:fix-pattern-distribution-framework} presents the distribution of fix patterns across libraries. While low-frequency fix patterns (AEH, AID, and CPO) appear in only one library, high-frequency patterns like Change Version (CV) and Change References (CR) occur predominantly in LangChain across all data sources. This is unsurprising given LangChain's rapid release cycle, with over 45 new versions in 2025 and over 55 versions in 2024 \cite{langchain-pypi}. The most common fixes involve pinning specific library versions in dependency files and updating import statements when modules shift between packages across versions. Developers also frequently need to replace deprecated API calls with their updated equivalents based on migration guides. Additionally, data access fixes (FDA), though less common, appear primarily in custom agent implementations. Developers building custom agents often need to add parsing logic to extract structured data from LLM responses or implement error handling to access nested attributes in API responses, tasks that library-based implementations handle automatically through built-in parsers and response objects.

\vspace{-5pt}
\begin{mdframed}[backgroundcolor=gray!5]
Fix patterns are skewed, with version and reference changes dominating LangChain and data access fixes in custom agents.
\end{mdframed}
\vspace{-.3cm} 
\subsubsection{Challenges:}
To understand the challenges developers face in fixing agentic code, we explored the bugs fixes and listed some of the major issues in this field. 

\textbf{Feature Gaps: }Since many parts of this field are still underdeveloped, agent frameworks and libraries often miss needed features such as custom filters, embedding support, or sophisticated integration with external tools. Frameworks' reliance on pre-built modules makes it difficult to implement highly customized algorithms or complex data processing flows that forces developers to build outside the framework's abstraction. For instance, as of January 2026, LangChain supports basic indexing workflows, but retrieval logic is usually built by wiring components together. Developers therefore have to write custom tools, build wrappers, or extend base classes to meet their specific needs, which leads to a higher number of Addition of Operations–type fix patterns (e.g., \cite{stackoverflow79753835} and \cite{stackoverflow79497660}). There is also a significant gap between research and implementation, particularly in agent optimization. Although research has explored agent tuning methods \cite{zeng2024agenttuning}, these approaches are not integrated with mainstream libraries and consequently get neglected by practitioners. Achieving optimal agent performance requires systematic tuning of embedding models, LLMs, and chunking parameters, a computationally expensive process since SoTA LLMs contain billions of parameters. However, because frameworks lack built-in support for these optimization techniques, particularly cost-efficient tuning approaches, developers must either avoid optimization altogether or implement custom tuning solutions on their own.

\textbf{Version Instability: }Secondly, due to rapid changes in library versions, developers frequently encounter breaking changes to functionality or API references that disrupt existing implementations and require code modifications to restore compatibility (e.g. \cite{stackoverflow76726419} and \cite{stackoverflow76313568}). These changes often cause agent crashes or introduce new bugs that were not present in prior versions, which is reflected in the increase of the change version type fix patterns. Consequently, developers must navigate a tradeoff between using newer versions with additional features and maintaining stable implementations with older, more reliable versions.

\textbf{Opaque Behavior: }Lastly, the black box nature of LLMs makes certain bugs difficult to debug or reproduce. For instance, in a Stack Overflow post \cite{Aapolaris2025langchain_memory}, an agent fails to answer questions about previous conversations, which suggests a memory issue. However, further investigation reveals that the problem stems from how the LLM interprets different prompts rather than an actual memory failure. The question `\textit{What's my name?}' fails to retrieve information from chat history, while a rephrased version `\textit{Do you know my name?}' successfully accesses the same information. This behavior varies across different LLM providers, making it difficult for developers to diagnose and localize whether issues originate from the framework, the model, memory, or the prompt design. In this situation, applying fixes to an LLM agent becomes challenging.

\begin{mdframed}[backgroundcolor=gray!5]Developers struggle to fix AI agent code because frameworks lack needed features, libraries change too frequently breaking existing code, and AI models behave unpredictably making bugs hard to diagnose.
\end{mdframed}

As developers continue to face these challenges, the situation becomes worse when issues fail to produce correct output without any error logs. In this study, to assist developers, we build a multi-agent solution that automatically fixes bugs in LLM agents. The solution is described in the Section \ref{sec:methodology}.

\begin{figure*}[h]
    \centering
    \includegraphics[height=0.35\textheight]{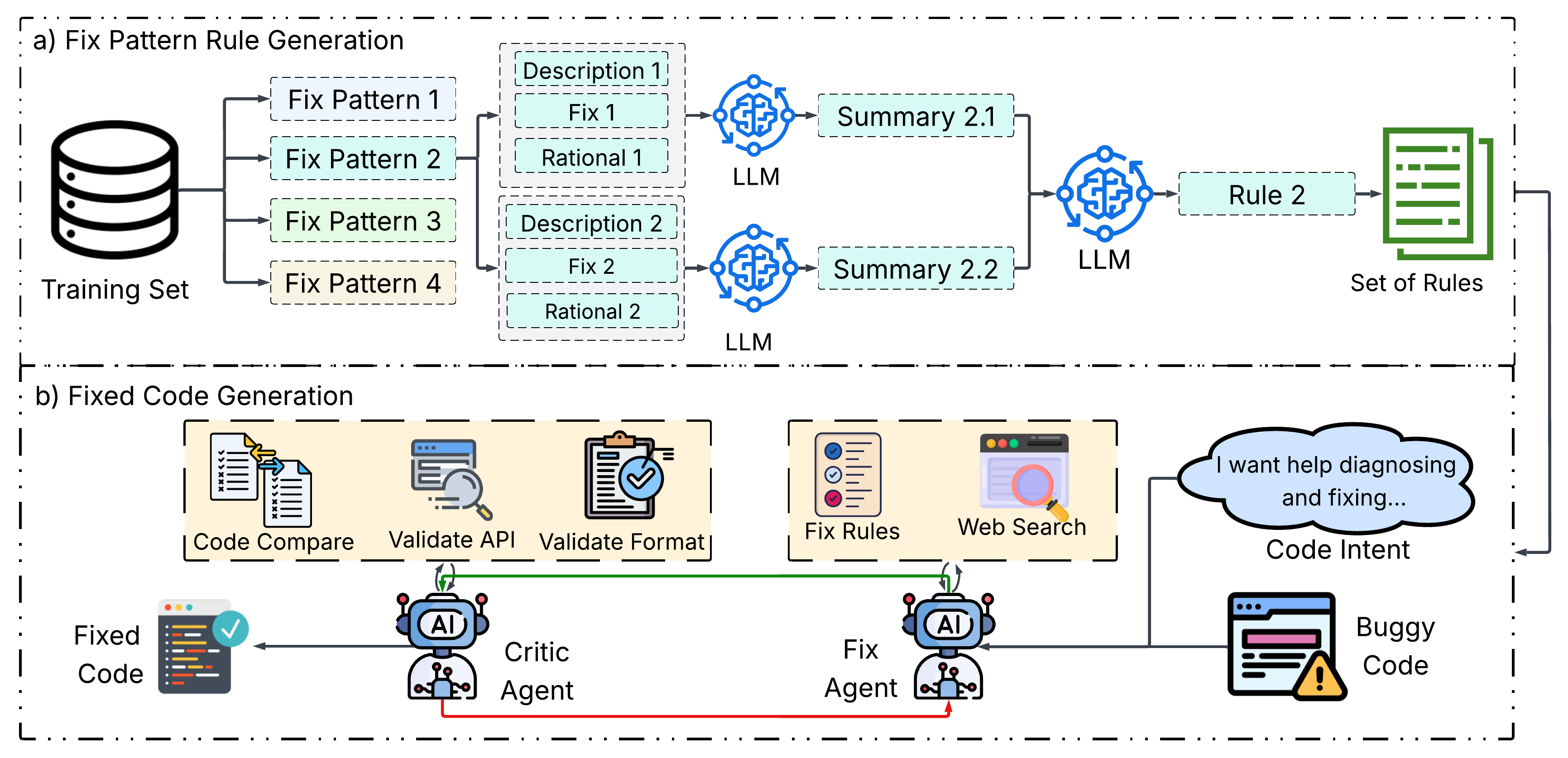}
    \vspace{2pt}
    \caption{Workflow of the proposed multi-agent approach.}
    \label{fig:approach}
\end{figure*}
\vspace{-8pt}
\subsection{Implications}
The prevalence of specific fix patterns across support forums suggests a shifting paradigm in the maintenance of agentic software. 

The dominance of \textit{Change Parameter Value (CPV)} type fix patterns highlights a critical need for specialized mutation testing frameworks tailored for agentic systems. Traditional testing fails to account for the high sensitivity of stochastic parameters (e.g. \cite{stackoverflow79313470}). Although many research works have leveraged LLMs for conducting mutation testing \cite{tip2025llmorpheus}, few studies have explored applying mutation testing to LLM agents. However, agentic systems require high computational costs to operate. This makes large-scale testing expensive. There is also a gap between research and implementation in cost-efficient testing approaches. While mutation testing techniques exist, they have not been adapted for the resource constraints of agentic systems. Few studies have focused on developing cost-efficient testing mechanisms for agentic systems, indicating a research gap that warrants further investigation.

Similarly, the high frequency of \textit{Change External Resources (CER)} type fix patterns in HuggingFace Forums highlights the necessity of standardized protocols for integrating open-source LLMs. Each open-source LLM has different API structures, input formats, and deployment requirements. For example, models like Llama \cite{touvron2023llama}, Qwen \cite{bai2023qwen}, and DeepSeek \cite{guo2025deepseek} require different configuration approaches for hosting and inference. Switching between LLM providers requires minimal code changes only when using unified APIs, but multi-backend libraries often face dependency bloat. Additionally, switching between tools presents substantial integration challenges. Although some solutions exist to address this issue \cite{ding2025unified}, they have not been widely adopted that forces developers to write custom integration code for each model and tool.

The prevalence of \textit{Change Version} fixes highlights the need for intelligent conflict resolver tools that identify dependency mismatches and recommend verified versions. While version migration tools exist for traditional software, they lack support for agent-specific breaking changes in memory management, tool execution, and prompt handling, or for detecting LLM API interaction issues. This forces manual conflict resolution, making automated solutions for agentic systems a critical research gap.

\vspace{-4pt}

\section{Approach}
\label{sec:methodology}
Figure \ref{fig:approach} provides an overview of the proposed solution, \name, which consists of two main steps. In the first step, fix pattern rules are extracted and subsequently used in the second step to generate fixes. In this context, a rule is a concise natural language description that specifies the conditions under which a particular fix pattern should be applied and how the code should be modified accordingly. The following sections describe these steps in detail.
\vspace{-6pt}
\subsection{Fix Pattern Rule Generation}
\label{sec:rule-generation}
The rule generation process comprises two main stages, as depicted in Figure \ref{fig:approach}. During the first stage, we generate individual summaries for each post associated with a given fix pattern. As shown in the diagram, for fix pattern 2, we process all associated Stack Overflow posts (two examples are shown) using an LLM. The summarization process takes as input the problem description (comprising the Stack Overflow post's title and body), the applied fix (either the corrected code or updated version), and the rationale provided by human annotators during the data annotation phase. The LLM processes this input to generate a concise one-line summary capturing both the problem and its solution. This process is repeated for every fix pattern in our taxonomy. Given the limited context window of LLMs, these summaries enable us to aggregate information from all posts associated with a fix pattern, a constraint that would otherwise prevent the inclusion of complete posts in the subsequent stage.
In the second stage, for each fix pattern, we collect all individual summaries and perform a comprehensive analysis to generate a unified rule. This rule encapsulates the general occurrence patterns of bugs and the corresponding fix strategies. Figure \ref{fig:approach} illustrates this two-stage process for fix pattern 2. The same procedure is applied across all fix patterns to generate the complete set of rules that guide the agent in subsequent steps. The generated rules provide a summary of each fix pattern and describe the issues in the category as well as the methods to apply to fix the issues. For rule generation, we used GPT 5 Mini, as it ranked as the best model for summarization according to the benchmark as of January 2026 \cite{prollm_summarization}.
\vspace{-10pt}

\subsection{Fixed Code Generation}
The proposed~\name  agent consists of two independent ReAct agents \cite{yao2023react} as the core component. A ReAct agent is an LLM-based agent that combines reasoning with actions to solve tasks. It operates through an iterative cycle that alternates between reasoning and action. The agent begins by reasoning about the problem to determine the optimal next step, then executes actions (such as tool invocations) that generate observations. These observations inform the subsequent reasoning step that creates a feedback loop that continues until the task is resolved. In the proposed solution, the first agent, the fix agent, generates a fix, while the second agent, the critic agent, evaluates the fix produced by the first agent.

The approach begins by passing the complete buggy code, along with the code intent generated in Step \ref{sec:benchmark-dataset}, to the fix agent. The agent analyzes the buggy code and attempts to create a corrected version using specific tools at its disposal. Details of the tools available to the fix agent are presented in Section \ref{sec:tools-fix-agent}.
\vspace{-8pt}
\subsubsection{Fix Agent Tools}
\label{sec:tools-fix-agent}
The fix agent has access to four tools that support rule lookup and external information retrieval during the fix generation process.

\textbf{List Fix Patterns:} This tool returns the names of all available fix patterns when the fix agent needs to address a problem by following the fix rules. Our study identified twenty-three fix patterns for LLM agents. If we include all pattern rules directly in the prompt, it would drastically increase token usage and costs. The agent may not need all these rules, so we provide only pattern names first through this tool. The agent can then selectively retrieve specific rules by passing the fix pattern name through the Fix Pattern Rule tool when it decides they are needed.

\textbf{Fix Pattern Rule:} After the agent receives the pattern names from List Fix Patterns, this tool (Fix Pattern Rule) retrieves the complete rule for a specific pattern. The rule explains what causes the issue and how to fix it. The agent analyzes this rule and then decides whether to modify code, request another pattern's rule, or continue without fixes. This two-step approach first lists names (through List Fix Patterns) and then fetches rules on demand (through Fix Pattern Rule). It minimizes unnecessary token consumption while it provides targeted guidance when needed.

\textbf{Web Search:} This tool allows the agent to search the internet for specific information using the Google Search API to retrieve relevant results. Since the evaluation dataset was collected from Stack Overflow and HuggingFace Forums, the tool filters out results from the corresponding source site (i.e., Stack Overflow results are excluded for Stack Overflow dataset instances, and HuggingFace results are excluded for HuggingFace dataset instances) to prevent potential data leakage.

\textbf{Submit Fix Code:} This tool is invoked when the fix agent has successfully generated a corrected version of the code. It saves the fixed code, which is later used by the critic agent for evaluation and stored in the file if it passes the agent's assessment.

Once the fix agent generates a complete fixed code, the critic agent is invoked with the original buggy code and the test code (generated by the authors in Stage \ref{sec:benchmark-dataset}). Following the fix agent, the critic agent has access to specific tools that it can invoke. The tools available to the critic agent are presented in Section \ref{sec:tools-critic-agent}.

\vspace{-7pt}
\subsubsection{Critic Agent Tools}
\label{sec:tools-critic-agent}

The critic agent has access to three specific tools for its decision-making process.

\textbf{Code Compare:} This tool compares the differences between the buggy code and the fixed code generated by the fix agent in the previous step. It uses the Python library difflib \cite{python-difflib} to identify and explain the changes, which the critic agent then evaluates in subsequent steps.

\textbf{Validate API:} Given the constantly evolving nature of the agentic field, validating the current functionality of APIs is essential. This tool uses web search to verify the usage of specific functions, methods, or API components. It searches for available parameter and return type documentation online and confirms whether the API is correctly invoked in the appropriate context. Like the fix agent, this tool excludes the source website from search results to prevent data leakage.

\textbf{Validate Format:} This tool ensures the fixed code format aligns with the buggy code format so the test file can execute properly. For instance, if the test file requires a function named \texttt{run\_agent}, this tool verifies that the code includes the correct formatting and structure, such as ensuring the agent returns output through a function rather than printing it directly. It also ensures the fixed code can be executed on the test file without any modifications.

With the help of the available tools, the critic agent generates a verdict (accept or reject) along with supporting reasoning. If the critic agent decides to reject, the fix agent analyzes the reasoning and generates another version of the fixed code, which is then passed back to the critic agent. This iterative loop between the two agents continues for a maximum of three iterations. If the critic agent rejects the third version of the fixed code, the last code generated by the fix agent is returned as the final output.

\vspace{-2pt}
\vspace{-6pt}
\section{Evaluation}
\label{sec:results}

We analyze our proposed solution to answer the following research question.
\vspace{-5pt}
\begin{itemize}
    \item \textbf{RQ5 (Effectiveness):} How effective is our proposed approach at localizing and fixing bugs in agentic systems?
    \item \textbf{RQ6 (Cost):} What is the average cost and time required by our solution?
    \item \textbf{RQ7 (Ablation):} What is the individual impact of fix-pattern\allowbreak–extracted rules and the web search on the effectiveness of the proposed system? 
    \item \textbf{RQ8 (Comparison):} How effective is our approach compared to existing solutions? 
\end{itemize}
\vspace{-8pt}
\subsection{Experimental Setup}
The experiment was conducted on a MacBook Air with an M3 processor and 16 GB RAM. For invoking LLMs, we leveraged the OpenAI API (for GPT-5.2) and OpenRouter API (for Gemini 3 Pro and Claude Sonnet 4). For web search, we used SerpAPI for Google Search. Lastly, we used LangChain \cite{mavroudis2024langchain} and LangGraph \cite{langgraph-official} with Python programming language for building the solution. 
\vspace{-2mm}
\subsection{Real-World Benchmark Dataset}
\label{sec:benchmark-dataset}
While our empirical analysis, presented in Section \ref{sec:empirical_study}, focused mainly on fix patterns, the collected dataset also includes buggy code, fixed code, error messages for crash bugs, and output logs for bugs producing wrong output. To better understand the nature of failures in real world settings, we further analyze the dataset. Our analysis on the Stack Overflow dataset shows that although crash bugs occur over ten times more frequently than bugs producing wrong output, the latter are significantly more difficult to debug. Many crash bugs can be debugged by analyzing the crash report, which helps developers identify the failure point \cite{medeiros2024impact, xu2023method}. Additionally, some LLM-powered solutions have been utilized for automatically fixing crash bugs in software systems \cite{huang2025one}. By contrast, wrong or missing output provides little diagnostic information that makes it ambiguous which component caused the failure.

To address this concern, we present the first benchmark dataset, \benchmark, containing bugs in LLM agents. We collected buggy posts from Stack Overflow and HuggingFace forums that generate incorrect output. Since these posts generally contain partial code focused on the buggy segment, we added minimal additional code while preserving the user-provided segment intact to reproduce the bug. We also generated a fixed version of the code following the accepted answer and created a test file to validate the fix. We further provided a markdown file explaining how to run the code and a requirements file containing the necessary libraries and their appropriate versions. Since all code segments are executable and produce no error messages, additional context about the user's intention is required. To address this, we generated a one-line code intent describing what the user expects the code to perform by passing the post's title and body to an LLM (GPT 5 Mini), which was then verified by a human. We selected GPT 5 Mini due to its performance in summarization tasks \cite{prollm_summarization}.

The collected \benchmark dataset consists of bugs found in the two most popular libraries, namely LangChain \cite{mavroudis2024langchain} and LlamaIndex \cite{llamaindex}, as these two libraries account for 60.1\% of the total bugs found in the Stack Overflow dataset. In total, we collected 37 code entries, with 36 sourced from Stack Overflow and 1 from HuggingFace forums. Additionally, we ensured that the component where the bug occurred spans across all four components of LLM agents, with a maximum of 19 out of 37 bugs occurring in the tools. In summary, each buggy code entry includes the buggy code, fixed code, test code, a markdown file with execution instructions, a requirements file listing the prerequisite libraries to install, and a code intent explaining what the user wants the code to perform.
\vspace{-6pt}
\subsection{Result Analysis}

\subsubsection{Effectiveness}
To evaluate \name, we tested it on the newly curated \benchmark dataset. Following prior work \cite{xia2025demystifying}, we report the resolution rate and localization accuracy at three levels of granularity: line, function, and LLM agent component. We selected three LLMs for evaluation: Claude Sonnet 4, Gemini 3 Pro, and GPT 5.2, representing the best-performing models for agentic coding as of December 2025 across three providers, based on SWE Benchmark results \cite{jimenez2023swe} \cite{vellum_llm_leaderboard_2025}. Table~\ref{tab:individual-comparison-table} presents the results. \name equipped with Gemini 3 Pro achieved the highest resolution rate by correctly fixing 22 out of 37 buggy code segments, followed by Claude Sonnet 4 with 21 and GPT-5.2 with 20. Localization followed a similar pattern: while Gemini 3 Pro and Claude Sonnet 4 tied on line-level localization, Gemini 3 Pro led in both function and component levels. GPT 5.2 scored the lowest overall. 
\vspace{-4pt}
\begin{mdframed}[backgroundcolor=gray!5]
 \name with Gemini 3 Pro achieved the highest performance by solving 22 of 37 bugs in the \benchmark dataset.
\end{mdframed}
\vspace{-15pt}
\subsubsection{Cost}
\begin{figure*}[htbp]
    \centering
   
    \begin{subfigure}[b]{0.48\textwidth}
        \centering
        \includegraphics[width=\linewidth]{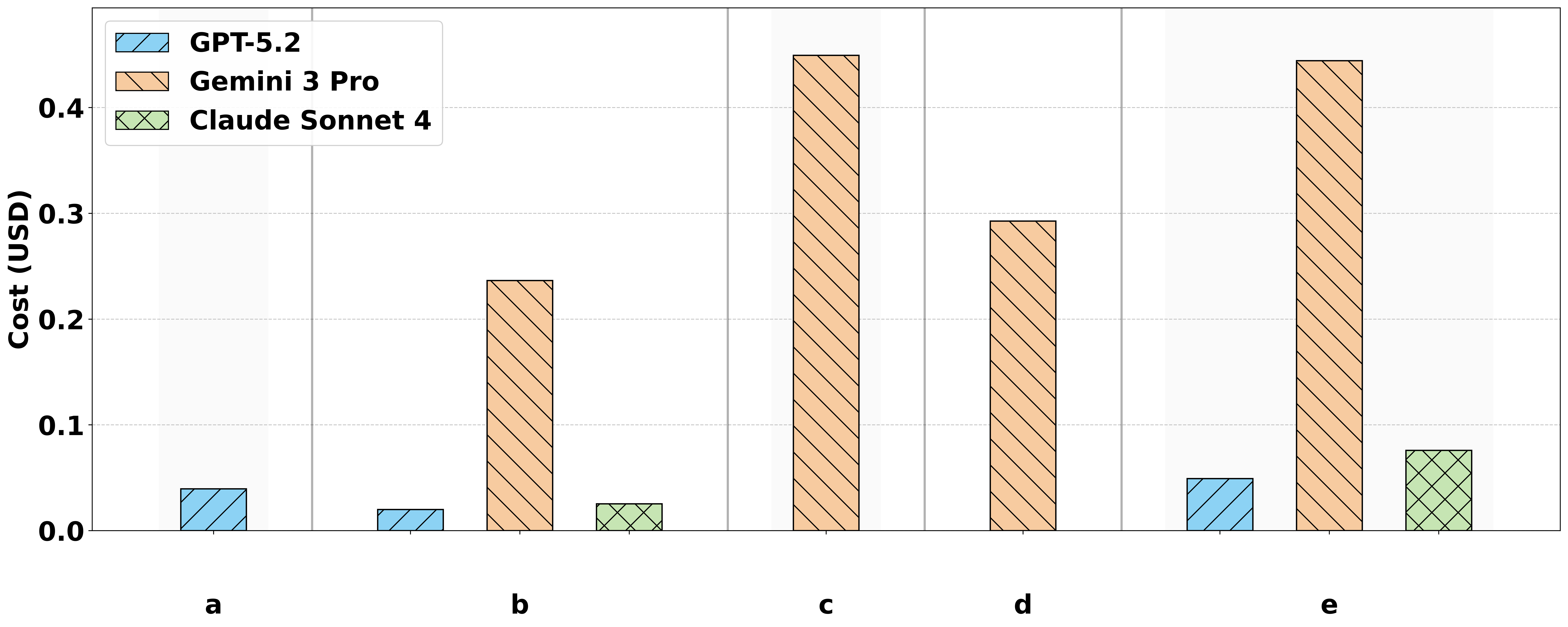}
        \caption{Average cost comparison}
        \label{fig:cost_comparison}
    \end{subfigure}
    \hfill 
    \begin{subfigure}[b]{0.48\textwidth}
        \centering
        \includegraphics[width=\linewidth]{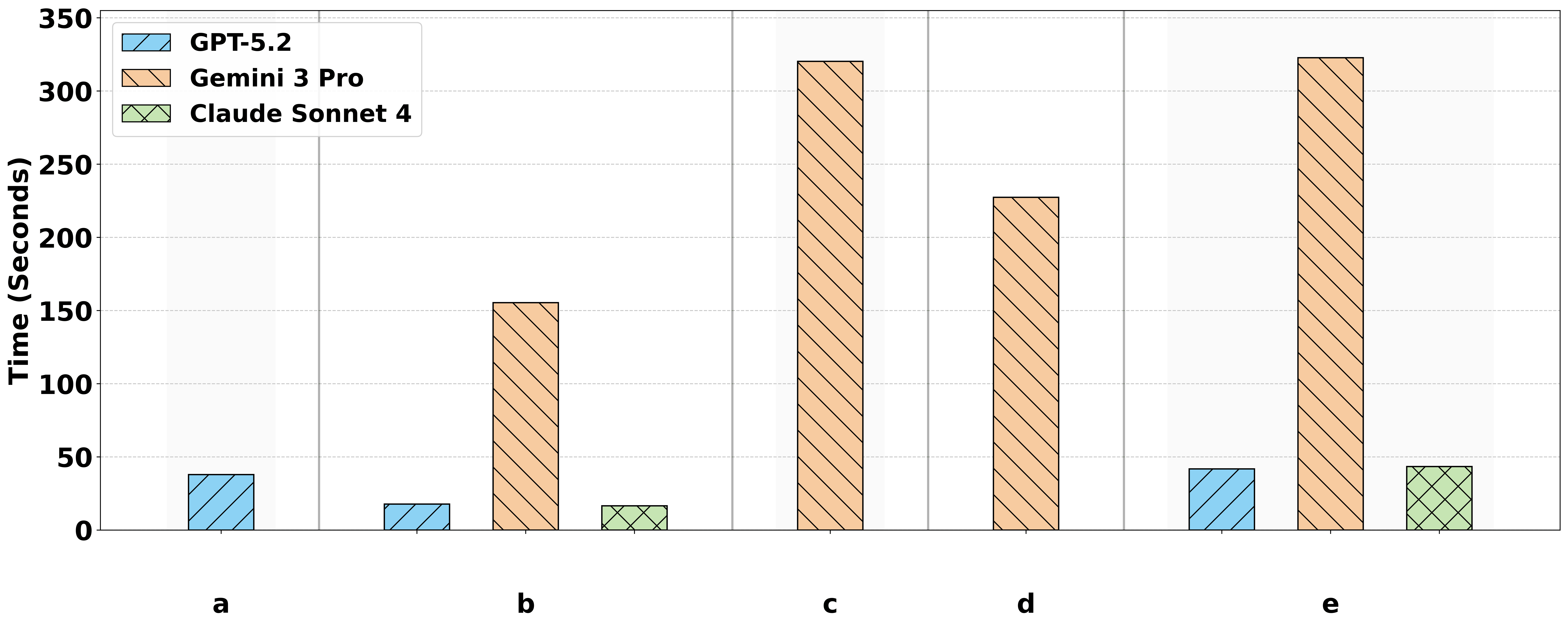}
        \caption{Average time comparison}
        \label{fig:time_comparison}
    \end{subfigure}
    
    \caption{Performance comparison of different approaches, including (a) SWE Agent, (b) Zero-shot, (c) No Fix Rules, (d) No Web Search, and (e) SelfHeal.}
    \label{fig:performance_comparison}
\end{figure*}

Figure \ref{fig:performance_comparison} presents a comparison of time and cost consumption across different LLMs and approaches. The diagram shows an inverse correlation between the performance and both time and cost. In \name, Gemini 3 Pro, with the highest performance, consumes an average of 322.72 seconds before reaching a solution, while the worst performer, GPT-5.2, consumes the lowest time of just 41.77 seconds. Similarly, Gemini 3 Pro consumes an average of 0.4442 USD for fixing a buggy code segment, while GPT-5.2 takes 0.0492 USD. Claude Sonnet 4 occupies a middle position with respect to both performance and cost. On average, the model incurs a cost of 0.0759 USD per buggy code segment and requires 43.40 seconds to generate a fix.

\begin{table*}[t]
\centering
\resizebox{\textwidth}{!}{
\begin{threeparttable}
\caption{Comparison between the proposed approach, baseline, and SoTA. Complete comparison shared in \cite{selfheal2026}}
\label{tab:individual-comparison-table}

\begin{tabular}{|c|cccccccccccc|cccccccccccc|cccc|}
\hline
\multirow{3}{*}{Post ID} & \multicolumn{12}{c|}{\name} & \multicolumn{12}{c|}{Zero-Shot} & \multicolumn{4}{c|}{SWE-Agent} \\ \cline{2-29} 
 & \multicolumn{4}{c|}{GPT-5.2} & \multicolumn{4}{c|}{Gemini 3 Pro} & \multicolumn{4}{c|}{Claude Sonnet 4} & \multicolumn{4}{c|}{GPT-5.2} & \multicolumn{4}{c|}{Gemini 3 Pro} & \multicolumn{4}{c|}{Claude Sonnet 4} & \multicolumn{4}{c|}{GPT-5.2} \\ \cline{2-29} 
 & \multicolumn{1}{c|}{RP} & \multicolumn{1}{c|}{LI} & \multicolumn{1}{c|}{FN} & \multicolumn{1}{c|}{CP} & \multicolumn{1}{c|}{RP} & \multicolumn{1}{c|}{LI} & \multicolumn{1}{c|}{FN} & \multicolumn{1}{c|}{CP} & \multicolumn{1}{c|}{RP} & \multicolumn{1}{c|}{LI} & \multicolumn{1}{c|}{FN} & CP & \multicolumn{1}{c|}{RP} & \multicolumn{1}{c|}{LI} & \multicolumn{1}{c|}{FN} & \multicolumn{1}{c|}{CP} & \multicolumn{1}{c|}{RP} & \multicolumn{1}{c|}{LI} & \multicolumn{1}{c|}{FN} & \multicolumn{1}{c|}{CP} & \multicolumn{1}{c|}{RP} & \multicolumn{1}{c|}{LI} & \multicolumn{1}{c|}{FN} & CP & \multicolumn{1}{c|}{RP} & \multicolumn{1}{c|}{LI} & \multicolumn{1}{c|}{FN} & CP \\ \hline \hline
76906469 & \multicolumn{1}{c|}{\checkmark} & \multicolumn{1}{c|}{\checkmark} & \multicolumn{1}{c|}{\checkmark} & \multicolumn{1}{c|}{\checkmark} & \multicolumn{1}{c|}{\checkmark} & \multicolumn{1}{c|}{\checkmark} & \multicolumn{1}{c|}{\checkmark} & \multicolumn{1}{c|}{\checkmark} & \multicolumn{1}{c|}{\checkmark} & \multicolumn{1}{c|}{\checkmark} & \multicolumn{1}{c|}{\checkmark} & \checkmark & \multicolumn{1}{c|}{$\times$} & \multicolumn{1}{c|}{$\times$} & \multicolumn{1}{c|}{$\times$} & \multicolumn{1}{c|}{$\times$} & \multicolumn{1}{c|}{$\times$} & \multicolumn{1}{c|}{\checkmark} & \multicolumn{1}{c|}{\checkmark} & \multicolumn{1}{c|}{\checkmark} & \multicolumn{1}{c|}{\checkmark} & \multicolumn{1}{c|}{\checkmark} & \multicolumn{1}{c|}{\checkmark} & \checkmark & \multicolumn{1}{c|}{$\times$} & \multicolumn{1}{c|}{$\times$} & \multicolumn{1}{c|}{\checkmark} & \checkmark \\ \hline
79753835 & \multicolumn{1}{c|}{$\times$} & \multicolumn{1}{c|}{$\times$} & \multicolumn{1}{c|}{$\times$} & \multicolumn{1}{c|}{\checkmark} & \multicolumn{1}{c|}{\checkmark} & \multicolumn{1}{c|}{\checkmark} & \multicolumn{1}{c|}{\checkmark} & \multicolumn{1}{c|}{\checkmark} & \multicolumn{1}{c|}{$\times$} & \multicolumn{1}{c|}{\checkmark} & \multicolumn{1}{c|}{\checkmark} & \checkmark & \multicolumn{1}{c|}{$\times$} & \multicolumn{1}{c|}{$\times$} & \multicolumn{1}{c|}{$\times$} & \multicolumn{1}{c|}{\checkmark} & \multicolumn{1}{c|}{$\times$} & \multicolumn{1}{c|}{$\times$} & \multicolumn{1}{c|}{\checkmark} & \multicolumn{1}{c|}{\checkmark} & \multicolumn{1}{c|}{$\times$} & \multicolumn{1}{c|}{\checkmark} & \multicolumn{1}{c|}{\checkmark} & \checkmark & \multicolumn{1}{c|}{$\times$} & \multicolumn{1}{c|}{$\times$} & \multicolumn{1}{c|}{$\times$} & $\times$ \\ \hline
79497660 & \multicolumn{1}{c|}{$\times$} & \multicolumn{1}{c|}{$\times$} & \multicolumn{1}{c|}{$\times$} & \multicolumn{1}{c|}{\checkmark} & \multicolumn{1}{c|}{\checkmark} & \multicolumn{1}{c|}{\checkmark} & \multicolumn{1}{c|}{\checkmark} & \multicolumn{1}{c|}{\checkmark} & \multicolumn{1}{c|}{\checkmark} & \multicolumn{1}{c|}{\checkmark} & \multicolumn{1}{c|}{\checkmark} & \checkmark & \multicolumn{1}{c|}{$\times$} & \multicolumn{1}{c|}{$\times$} & \multicolumn{1}{c|}{$\times$} & \multicolumn{1}{c|}{\checkmark} & \multicolumn{1}{c|}{$\times$} & \multicolumn{1}{c|}{$\times$} & \multicolumn{1}{c|}{\checkmark} & \multicolumn{1}{c|}{\checkmark} & \multicolumn{1}{c|}{$\times$} & \multicolumn{1}{c|}{$\times$} & \multicolumn{1}{c|}{\checkmark} & \checkmark & \multicolumn{1}{c|}{$\times$} & \multicolumn{1}{c|}{$\times$} & \multicolumn{1}{c|}{$\times$} & \checkmark \\ \hline \hline
Total & \multicolumn{1}{c|}{20} & \multicolumn{1}{c|}{24} & \multicolumn{1}{c|}{29} & \multicolumn{1}{c|}{33} & \multicolumn{1}{c|}{22} & \multicolumn{1}{c|}{27} & \multicolumn{1}{c|}{31} & \multicolumn{1}{c|}{34} & \multicolumn{1}{c|}{21} & \multicolumn{1}{c|}{27} & \multicolumn{1}{c|}{29} & 33 & \multicolumn{1}{c|}{12} & \multicolumn{1}{c|}{18} & \multicolumn{1}{c|}{22} & \multicolumn{1}{c|}{24} & \multicolumn{1}{c|}{13} & \multicolumn{1}{c|}{19} & \multicolumn{1}{c|}{29} & \multicolumn{1}{c|}{30} & \multicolumn{1}{c|}{15} & \multicolumn{1}{c|}{22} & \multicolumn{1}{c|}{29} & 32 & \multicolumn{1}{c|}{14} & \multicolumn{1}{c|}{18} & \multicolumn{1}{c|}{27} & 30 \\ \hline
\end{tabular}

\begin{tablenotes}
\small
\item \textbf{Abbreviations:} \textbf{RP}: Repair; \textbf{LI}: Line Identification; \textbf{FN}: Function identification; \textbf{CP}: Component identification; \textbf{\name}: Proposed Approach.
\end{tablenotes}
\end{threeparttable}}
\end{table*}

\vspace{-3pt}

\begin{mdframed}[backgroundcolor=gray!5]
There is an inverse correlation between performance and both time and cost. The highest-performing model consumes the most time and incurs the highest cost.
\end{mdframed}
\vspace{-17pt}
\subsubsection{Ablation}
To measure the contribution of each component, we evaluated our solution by removing two data sources: internal and external. First, to assess the contribution of the internal data source, we removed the fix rules (fix pattern list and fix pattern rule tool) from the fix agent, allowing it to rely only on the internet search for external data. This resulted in an 18.92\% performance drop. Next, we removed the web search tool from the fix agent and the validate API tool from the critic agent (which uses web search in the backend), resulting in a 13.51\% performance drop. These results demonstrate the model's reliance on both internal and external data sources. Furthermore, in three instances, the fix agent generated incorrect fixes that were detected by the critic agent, which then led to successful corrections. Table \ref{tab:results-compare} illustrates the performance of different models and settings in detail. Since Gemini 3 Pro produced the best result, the ablation study was conducted using that LLM.
\vspace{2pt}
\begin{mdframed}[backgroundcolor=gray!5]
Removing internal data sources degrades the repair performance by 18.92\%, compared to 13.51\% for external sources.
\end{mdframed}
\vspace{19pt}

\subsubsection{Comparison}
\vspace{-5pt}
\begin{table}[ht]
\centering
\begin{threeparttable}
\caption{Performance comparison of bug identification and repair using different approaches.}
\label{tab:results-compare}

\begin{tabular}{|
>{\centering\arraybackslash}p{0.135\linewidth} |
>{\centering\arraybackslash}p{0.05\linewidth} |
>{\centering\arraybackslash}p{0.1\linewidth} |
>{\centering\arraybackslash}p{0.08\linewidth} |
>{\centering\arraybackslash}p{0.1\linewidth} |
>{\centering\arraybackslash}p{0.11\linewidth} |
>{\centering\arraybackslash}p{0.10\linewidth} |}
\hline
\textbf{App.} & \textbf{M.} & \textbf{Repair} & \textbf{Line} & \textbf{Fn.} & \textbf{Comp.} & \textbf{Attmp.} \\ \hline \hline
SWE-A & G5.2 & 0.3784 & 0.4865 & 0.7297 & 0.8108 & 3.78 \\ \hline
\multirow{3}{*}{Zero-shot} & G5.2 & 0.3243 & 0.4865 & 0.5946 & 0.6486 & 1.00 \\ \cline{2-7}
 & G3P & 0.3514 & 0.5135 & 0.7838 & 0.8108 & 1.00 \\ \cline{2-7}
 & CS4 & 0.4054 & 0.5946 & 0.7838 & 0.8649 & 1.00 \\ \hline
NFR & G3P & 0.4054 & 0.5676 & 0.7568 & 0.8919 & 1.30 \\ \hline
NWS & G3P & 0.4595 & 0.5135 & 0.8378 & 0.8649 & 1.24 \\ \hline
NCA & G3P & 0.5135 & 0.7297 & 0.8378 & 0.9189 & 1.32 \\ \hline
\multirow{3}{*}{\name} & G5.2 & 0.5405 & 0.6486 & 0.7838 & 0.8919 & 1.35 \\ \cline{2-7}
 & G3P & 0.5946 & 0.7297 & 0.8378 & 0.9189 & 1.32 \\ \cline{2-7}
 & CS4 & 0.5676 & 0.7297 & 0.7838 & 0.8919 & 1.05 \\ \hline
\end{tabular}

\begin{tablenotes}
\small
\item \textbf{App.}: Approach; \textbf{M.}: Model; \textbf{G5.2}: GPT-5.2; \textbf{G3P}: Gemini 3 Pro; \textbf{CS4}: Claude Sonnet 4; \textbf{Fn.}: Function; \textbf{Comp.}: Component; \textbf{Attmp.}: attempts.; 
\textbf{SWE-A}: SWE-Agent;
\textbf{NFR}: No fix rules; \textbf{NWS}: No Web Search; \textbf{NCA}: No Critic Agent.
\end{tablenotes}
\end{threeparttable}

\end{table}
\vspace{-1cm}
We compared our approach against a zero-shot baseline and the SoTA SWE-Agent \cite{yang2024swe}, which has been shown to outperform prior methods on software engineering benchmarks. Although the original SWE-Agent study leveraged GPT-4 Turbo and Claude 3 Opus, we adopt GPT-5.2 as the base model to ensure a fair comparison. Table~\ref{tab:results-compare} presents a detailed comparison of the approaches. The results show that the proposed method significantly outperforms both the baseline and SWE-Agent, with an average repair rate improvement of 21.72\% over the baseline and 16.21\% over SWE-Agent. Although Gemini 3 Pro yields the best performance in \name, Claude Sonnet 4 attains the highest repair rate among zero-shot approaches. This suggests that the performance gains of Claude Sonnet 4 stem mainly from the strength of the base model rather than external tools when compared with the three LLMs used. This pattern is also reflected in the number of attempts, as the LLM often produces a correct fix on the first attempt and needs fewer attempts afterward. In contrast, SWE-Agent requires more attempts on average to reach a solution.
\begin{mdframed}[backgroundcolor=gray!5]
\name outperforms both baseline and SoTA approach by a significant margin. 
\end{mdframed}


\subsection{Discussion}
Overall, \name demonstrated noteworthy performance in both localizing and repairing buggy agentic code. This higher performance can be attributed to the careful design of the architecture. While existing agentic systems like SWE-Agent \cite{yang2024swe}, AutoCoderover \cite{zhang2024autocoderover}, and Agentless \cite{xia2025demystifying} also employ step-by-step reasoning and feedback processes, their integrated tools do not provide LLMs with additional information from the external environment. In rapidly developing fields like LLM agents, relying solely on an LLM's training knowledge is insufficient, as it quickly becomes obsolete. Therefore, our solution provides additional information through internal (fix-rule) and external (web search) sources, which enhances the agent's capabilities in conjunction with the critic agent that critiques the code and provides additional feedback. Nonetheless, this solution incurs a higher cost and time when using the best-performing LLM.

Although the proposed solution achieved noteworthy performance, outperforming both baseline and state-of-the-art approaches, it has several limitations that can be addressed in future studies. First, the \benchmark dataset contains only 37 instances, smaller than other benchmarks used in software engineering tasks such as the 40 instances in the buggy DNN code benchmark \cite{wardat2021deeplocalize}. This limited size is due to LLM-based agents for code debugging being a relatively new field, making collection of real-world buggy code that does not crash challenging. Additionally, it consists of bugs from only the two most popular libraries. Second, although the benchmark contains complete executable code, \name does not execute code during debugging to leverage dynamic features, instead relying on an internal dataset of fix rules and internet search. While executing code and extracting dynamic features would significantly increase solution time, future work can explore this trade-off.  Lastly, the solution has not been evaluated on repository-level codebases, unlike tools such as SWE-Agent \cite{yang2024swe}, AutoCoderover \cite{zhang2024autocoderover}, and requires modifications before adoption for fixing bugs in large-scale agentic systems with multiple files.
\vspace{-9pt}

\section{Related work}
\label{sec:related}

   This section reviews existing work on software bugs in deep learning, the evolution of Agentic AI in software engineering, and prior research on agent-based bug repair.
\vspace{-5pt}
\paragraph{\textbf{Study on Software bug:}}
Several studies investigated software bug types across traditional systems and modern deep learning applications. Early work by Catolino et al defined a taxonomy of common software bugs and introduced an automated classification model\cite{catolino2019not}. As machine learning systems became prevalent, comprehensive analyses showed data and logic bugs from incorrect parameters were most common\cite{islam2019comprehensive}, leading to a validated taxonomy based on practitioner interviews\cite{humbatova2020taxonomy}. Du \etal \cite{du2024llm} introduced a framework using language model embeddings to automatically classify deep learning bug reports. Jahan~\etal~\cite{jahan2025taxonomy} presented a taxonomy of faults in attention-based neural networks, showing many failures are unique to attention mechanisms. Han \etal \cite{han2025comprehensive} studied bugs in foundation language models and found that dependency and API issues are the main causes of crashes. Pan \etal \cite{pan2024lost} analyzed bugs from LLM-based code translation and identify translation-specific error categories. Yu \etal \cite{yu2025towards} analyzed 308 bugs in large scale LLM training systems and identify common causes, challenges in debugging. Islam \etal \cite{islam2026agentsfail} conducted the first large-scale study on bugs in LLM agents. Xue \etal \cite{XueCharacterization}, on the other hand, studied the bugs in LLM agent building frameworks. However, no study has analyzed bug-fix patterns specifically in agentic systems. Given the rapid development of LLM agents, we conducted the first study on bug fix patterns in this field.
\vspace{-7pt}
\paragraph{\textbf{Agentic AI and SE}}
Recent work has increasingly explored the role of Agentic AI in advancing software engineering practices.
Liu ~\etal~\cite{liu2024large} provided a comprehensive systematic survey on LLM-based agents in Software Engineering, categorizing them by both SE applications and agent architectures. Building on this foundation, Yang ~\etal~ \cite{yang2024swe} demonstrated that providing LLMs with a custom interface via SWE-agent leads to state-of-the-art results in autonomous code repair on the SWE-bench dataset. Zhang ~\etal~\cite{zhang2024mleagent} proposed MLE-Agent, an autonomous assistant that streamlines ML development through a combination of research tools and automated debugging. Beyond single-agent systems, He ~\etal~ \cite{he2025llm} examined the role of multi-agent LLM systems in software engineering and outlined a path toward scalable, autonomous development. Terragni ~\etal~ \cite{terragni2025future} explored the future of human-AI synergy in coding while highlighting the technical challenges ahead. Furthermore, Liang ~\etal~ \cite{liang2025can} presented the RepoCod benchmark to evaluate Python generation, demonstrating that current models often fail on full-scale software tasks.
\vspace{-7pt}
\paragraph{\textbf{Agent-Based bug Repair}} Recent studies have systematically examined agent-based approaches for automated bug repair. Deshpande ~\etal~ \cite{deshpande2025trail} introduced a dataset and error taxonomy for evaluating agentic workflow traces, showing that current LLMs failed significantly at automated debugging. Complementing this empirical analysis, Epperson ~\etal~ \cite{epperson2025interactive} introduced an interactive debugging tool that enables developers to visualize complex histories and reset agent messages to resolve common challenges. From a defect analysis perspective, Ning ~\etal~ \cite{ning2024defining} presented a systematic study and static analysis tool to detect agent defects by analyzing discrepancies between developer logic and LLM-generated content. They benchmarked real-world agent issues and found that SoTA LLM agents struggle to resolve agent-specific bugs. To address this gap, we curated the first benchmark on runtime bugs in LLM agents and developed \name, an LLM agent capable of fixing bugs in agentic systems.

Overall, this study addresses two major research gaps in LLM agents. First, the lack of empirical analysis on bug fix patterns in LLM agents, and second, the limited ability of SoTA LLM agents to fix bugs in agentic systems.
\vspace{-.1cm}

\vspace{-.3cm}
\section{Threats to Validity}
\label{sec:threats}

\textbf{Internal Validity:} A potential threat arises from the correctness of the annotated dataset, as fix patterns related to agents involve complex and stochastic behaviors. We mitigate this risk by employing two independent annotators with experience in agent development, measuring agreement with Cohen's Kappa coefficient, and resolving all conflicts through discussion with an expert. Another threat concerns the correctness of \name's implementation, which we address through independent code review by two authors and validation of experimental results. Finally, since the benchmark dataset was curated from publicly available sources and \name integrates web search, there is a risk of data leakage during internet searches. We mitigate this by restricting web search to the source website of each benchmark instance and verifying through ablation studies that the agent's fixing ability does not depend solely on external sources. We have also ensured the instances selected for the \benchmark dataset are excluded while generating the fix rules.

\textbf{External Validity:} A threat to generalizability stems from benchmark representativeness. We mitigate this by curating data from GitHub, Stack Overflow, and HuggingFace Forums, which cover diverse real-world agent deployments. Another threat relates to dataset validity, as unverified fixes may not reflect realistic scenarios. We ensure each instance includes an accepted answer, validated solution, or external reference. For externally sourced fixes, both annotators independently verified correctness and documented the references in the dataset. Finally, to reduce LLM stochasticity, we compare our approach against baselines under identical settings.

\vspace{-3mm}
\section{Conclusions and Future Work}
\label{sec:conc}
This study presents an empirical analysis of bug-fix patterns in LLM agents, comparing these patterns across different software domains, and provides actionable implications for the community. To facilitate further research on fixing bugs in LLM agents, we prepared an executable benchmark dataset,~\benchmark, containing runtime bugs with buggy code, fixed code, and corresponding test cases. We also built a multi-agent system,~\name, for automatically fixing bugs in LLM agents. Our analysis reveals that the proposed solution outperforms both the baseline and state-of-the-art approaches by a noticeable  margin. Despite these contributions, this study has several limitations that future research can address. First, the \benchmark dataset contains a relatively small number of instances and can be extended by exploring additional libraries beyond LangChain and LlamaIndex, as well as incorporating other data sources. Second, the proposed solution relies on static analysis for bug detection. Future studies can integrate dynamic analysis to determine whether runtime data improves bug-fixing performance. Finally, \name currently operates on single files containing buggy instances. Future work can extend the approach to handle repository-level codebases with multiple interdependent files.

\vspace{-5pt}
\section[Data Availability]{Data Availability}
\label{sec:data-availability}
The fix pattern dataset and benchmark dataset (\benchmark), along with results and code, are publicly available in the site \cite{selfheal2026} and GitHub \cite{selfheal2026-github}.

\bibliographystyle{ACM-Reference-Format}
\bibliography{sample-base}
\end{document}